\documentclass[final,3p,times,twocolumn]{elsarticle}
\usepackage{amsmath,amssymb,graphicx}
\usepackage{hyperref}
\usepackage{amsfonts,amsthm}

\usepackage{amsmath,amssymb,graphicx,amsthm}
\usepackage{color}
\begin{document}

\begin{frontmatter}

\title{Model reduction in chemical dynamics: slow invariant manifolds, singular perturbations, thermodynamic estimates, and analysis of reaction graph}
\author[LeicMath,NN]{A.N. Gorban}
 \ead{a.n.gorban@le.ac.uk}

\address[LeicMath]{Department of Mathematics, University of Leicester, Leicester, LE1 7RH, UK}
\address[NN]{Lobachevsky University, Nizhni Novgorod, Russia}

%\date{}

%\maketitle

\begin{abstract}

The paper has two goals:

\noindent(1) It presents basic ideas, notions, and methods for reduction of reaction kinetics models: quasi-steady-state, quasi-equilibrium, slow invariant manifolds, and limiting steps.

\noindent(2) It describes briefly the current state of the art and some latest achievements in the broad area of model reduction in chemical and biochemical kinetics, including new results in methods of invariant manifolds, computation singular perturbation,  bottleneck methods, asymptotology, tropical equilibration, and reaction mechanism skeletonisation.
\end{abstract}

\end{frontmatter}

\section*{Historical introduction} 

 Three eras (or waves) of chemical dynamics can be associated with their leaders \cite{GorbanYablonsky2015}: (1)  the van’t Hoff wave (the first Nobel Prize in Chemistry, 1901) (2)  the Semenov--Hinshelwood wave, and (3) the Aris wave. The problem of modelling of {\em  complex reaction networks} was in the focus of chemical dynamics research since the invention of the concept of  ``chain reactions'' by Semenov and Hinshelwood (the shared Nobel Prize in Chemistry, 1956). Aris' activity was concentrated on the detailed systematization of mathematical ideas and approaches for the needs of chemical engineering. In the engineering context, the problem of modelling of complex reactions became even more important.

 A mathematical model is an intellectual device that works \cite{GorbanYab2013}. Creation of such working models is impossible without the well developed technology of model reduction. Therefore, it is not surprising that the  model reduction methods were developed together with the first theories of complex chemical reactions. Three simple basic ideas have been invented:
 \begin{itemize}
 \item The {\em Quasi-Equilibrium} approximation or QE (a fraction of reactions approach their equilibrium fast enough and, after that, remain almost equilibrated);
 \item The {\em Quasi Steady State} approximation or QSS (some of species, very often these are some of intemediates or radicals, exist in relatively small amounts; they reach quickly their QSS concentrations, and then follow, as a slave, the dynamics of these other species remaining close to the QSS). The QSS is defined as  the steady state under condition that the concentrations of other species do not change;
 \item The {\em limiting steps} or bottleneck  is a relatively small part of the reaction network, in the simplest cases it is a single reaction, which rate  is a good approximation to the reaction rate of the whole network.
\end{itemize}
More precise formal discussion is  presented in the following sections. 
 
In   1980s --   1990s the model reduction technology was enriched by several ideas.  Most important of them are: the Method of Invariant Manifolds (MIM) theory and algorithms \cite{RouFra1990,  GKTTSP94}, the special Intrinsic Low Dimesional Manifold (ILDM) method for approximation of slow motion \cite{Maas1992},  the Computational Singular Perturbation (CSP) method for the iterative approximation of both slow and fast motions \cite{Lam1994}, and the sensitivity analysis of complex kinetic systems \cite{RabiSens1983}. 

Development of lumping analysis was important for general understanding of model reduction in chemical kinetics \cite{LiRabitz1990}. The lumped species is considered as a linear combination of the original ones. These combinations are often guessed on the basis of known kinetic properties and can be improved by iterative methods and observer theory. The standard examples are: (i) the lumped species are identified as the sums of species in selected groups (a very popular approach with many practical applications, for example, \cite{Ranzietal2008}); (ii) the lumped species are  the numbers of links and structural fragments of  various types and  in different states (this approach has many applications, from petrochemitry \cite{Nguyenetal2017} and modelling of intracellular networks \cite{Abou-Jaoude2016}  to the Internet dynamics \cite{Newman2011}).

The main achievements of this period (1980s --  1990s) in model reduction were summarised in several books and surveys \cite{GorbanKarlinBook2005, CMIM2004, TuranyiTomlin2014, GoussisMaas2011}.

The technological elaboration of these ideas  and assimilation of those by the modelling practice took almost thirty years. Much efforts have been invested into computational improvements and  testing with the systems of various complexity. Some new ideas were proposed and developed.

The QE, QSS, MIM, and CSP methods can be applied to any differential equation with explicit or implicit (hidden) separation of time. They use the structure of reaction network  as a tool for creation of kinetic equations. In the classical methods, only the limiting step approach (the bottleneck method) works directly with the reaction graph. Recently, the model reduction methods which use the structure of the reaction network, were developed far enough and attract many different techniques, from sensitivity analysis to algebraic geometry and tropical mathematics. 

The first step  in the next section is a ``step backwards'', a brief introduction of the classical methods. Then we move to modern development.
 
\section*{ QE, QSS, MIM and CSP  in ODE framework}

Formally, the standard models of chemical kinetics are systems of Ordinary Differential Equations (ODE). The general framework looks as follows. Let $U$ be a bounded domain in $R^n$. Assume that vector fields $F_{\rm fast}(x)$ and $F_{\rm slow}(x)$ are defined and differentiable in  a vicinity of $\bar{U}$  (in real applications these vector  fields are usually analytical, or even polynomial or rational). Let $U$ be positively invariant with respect to  $F_{\rm fast}(x)$ and $F_{\rm slow}(x)$. Consider dynamical system with the explicit {\em fast-slow time separation}:
\begin{equation}\label{EqSlowExpl}
\frac{d x }{d t}=F_{\rm slow}(x)+\frac{1}{\varepsilon}F_{\rm fast}(x),
\end{equation}
where $\varepsilon>0$ is a small parameter.
The {\em fast subsystem} is
\begin{equation}\label{EqFast}
\frac{d x }{d \tau}=F_{\rm fast}(x).
\end{equation}
Here, time $\tau$ is used to stress that this is the  `fast time', $\tau=t/\varepsilon$.
If the fast system (\ref{EqSlowExpl}) converges to an asymptotically stable fixed point in $U$ and has no fixed points on the border of $U$ then  for sufficiently small $\varepsilon$ the slow vector field becomes practically invisible in the dynamics of (\ref{EqSlowExpl}), i.e. there is no slow dynamics.

 Let the fast system (\ref{EqFast}) be neither globally stable nor ergodic in $U$. Assume that (i) it has the conservation laws $b_i(x)$ ($i=1,\ldots, k$) and (ii) for each $x_0\in U$ the fast system on the set $b_i(x)=b_i(x_0)$ converges to a unique stable fixed point   $x^*(b)$, where $b$ is the vector of values $b_i(x)$. Then the slow system describes dynamics of conservation laws $b$:
\begin{equation}\label{Eq:slow1}
\frac{d b}{d t}=Db_{x=x^*(b)}[F_{\rm slow}(x^*(b))],
\end{equation}
where $Db_{x=x^*(b)}$ is differential of $b(x)$ at the point $x^*(b)$. For linear conservation laws $Db=b$ and the slow equations have the simple form
\begin{equation}\label{Eq:simple}
\frac{d b}{d t}=b[F_{\rm slow}(x^*(b))].
\end{equation}
The {QE manifold} is parametrised by  the conservation laws with functions $x^*(b)$. It should be stressed that the slow equations in their natural form  (\ref{Eq:slow1}), (\ref{Eq:simple}) describe the dynamics of the conservation laws $b$ and {\rm not} the dynamics of the selected `slow coordinates'. The problem of projection onto slow manifold is widely discussed \cite{GearKaper2005, UNIMOLD2004, GorbanKarlin2003}. According to the  Tikhonov theorem, dynamics of the general system  (\ref{EqSlowExpl}) from an initial state $x_0$ under the given assumptions can be split in two stages: fast convergence to the QE manifold  $x^*(b)$ (the {\em initial layer}, convergence to a small vicinity of $x^*(x_0)$), and then slow motion in a small vicinity of the QE manifolds.

The QE assumption is the separation of reactions onto slow and fast: $F_{\rm slow}$ includes all the terms from the slow reactions, $F_{\rm fast}$ includes all the terms from the fast reactions and the slow manifold $x^*(b)$ consists of {\em equilibria} of the fast reactions parametrised by the conservation laws. The `thermodynamic' behaviour of fast reactions (convergence to  equilibrium, which is unique for any given values of the conservation laws) is  essential to application of the Tikhonov theorem. Slow reactions can be extended by including external fluxes, they do not change the asymptotic form (\ref{Eq:slow1}), (\ref{Eq:simple}). 

Combining of fast subsystems from the fast reactions is so popular \cite{LeeOthmer2010} that a special warning is needed: there exists another widely used approximation without separation of reactions into fast and slow (see QSS below).  

It should be stressed that the physical and chemical nature of the convergence to equilibrium of fast reactions may vary. It may follow from thermodynamic conditions like principle of detailed balance or semi-detailed balance. It may have also completely algebraic nature. For example, any linear (monomolecular or pseudomonomolecular) system tends to a fixed point, which is unique for any given values of conserved quantities. The mass action law systems with reactions `without interaction of different substances' $\alpha_{ri} A_i \to \sum_j \beta_{rj} A_j$ provide us with another example \cite{GorBykYab1986} (here $A_i$ are the components, $r$ is the number of reaction, and the coefficients are non-negative).

The QSS approximation does not separate the reactions into slow and fast. Usual application of the QSS is the slaving dynamics of the active intermediates in combustion (Semionov) or in catalysis (Hinshelwood). The stable reagents participate in the same reactions as the intermediates do and, therefore, we cannon consider them as `slow' reagents. The nature of small parameter in this case is very different \cite{SegelSlemrod1989}. It is the smallness of the {\em amount} of intermediates. Let us illustrate this small parameter for heterogeneous catalytic reactions \cite{Yab1991}. We use notations: $V$ is the volume of reactor, $S$ is the surface of catalyst, $N_{\rm g}$ is the vector of amounts of the gas components, $N_{\rm s}$ is the vector of amounts of the surface components,  $c_{\rm g}=N_{\rm g}/V$, $c_{\rm s} = N_{\rm s}/S$. Let us measure $V$ and $S$ in moles (i.e we use $PV/RT$ for some average values of the pressure $P$ and temperature $T$ instead of $V$ and the number of moles of active centres on the surface instead of $S$) for  creation of dimensionless variables. The kinetic equations are:
$$\frac{d N_{\rm g}}{dt}=SF_{\rm g}(c_{\rm g},c_{\rm s}), \;\frac{d N_{\rm s}}{dt}=SF_{\rm s}(c_{\rm g},c_{\rm s}),$$
Where $F_{\rm g}$ and $F_{\rm s}$ are combined from the reaction rated per unit surface (mole of active centres).
Notice that usually in the selected units $S\ll V$ (if the pressure is not extremely low). Under given $V$, we can write equations with a small parameter $\varepsilon=\frac{S}{V}$.
$$\frac{d c_{\rm g}}{dt}=\varepsilon F_{\rm g}(c_{\rm g},c_{\rm s}), \;\frac{d c_{\rm s}}{dt}=F_{\rm s}(c_{\rm g},c_{\rm s}).$$
After the change to the slow timescale, $\theta=\frac{S}{V}t$ we get the  slow and fast subsystems:
\begin{equation}\label{QSS2}
\dot{c}_{\rm g}= F_{\rm g}(c_{\rm g}, c_{\rm s}); \;\; \dot{c}_{\rm s}=\frac{1}{\varepsilon} F_{\rm s}(c_{\rm g},c_{\rm s}).
\end{equation}
If the fast dynamics of $c_{\rm s}$ tends to the asymptotically stable fixed point (quasi-steady-state) $c^*_{\rm s}(c_{\rm g})$ for any fixed value of $c_{\rm g}$ then we return to the previously described situation: the motion can be separated into two steps: first, $c_{\rm s}$ goes fast into a small vicinity of $c^*_{\rm s}(c_{\rm g})$ (initial layer) and then $c_{\rm g}$ slowly changes in vicinity of this manifold.

In the QSS approximation, the smallness of amount of intermediates is the crucial assumption. There may be various explanation of this smallness. For example, in catalysis this is the smallness of the amount of the catalyst. In combustion, the high activity of the intermediates leads to their short life time. 

The list of variables for QSS exclusion can change in time. The smallness of concentration can serve as a criterion for the extension/reduction of this list \cite{GorbanKarlin2003}. The `speed coefficients' were introduced recently \cite{WestBiktashev2015} for ranking variables according to how quickly their approach their momentary steady-state. These coefficients are used for  a straightforward choice of variables for QSS elimination , while preserving dynamic characteristics of the system.

In addition to the different nature of small parameters, the QSS differs from the QE in one more  essential aspect: for the QE, the fast system is just chemical kinetics of the set of fast reactions. It  has all the thermodynamic properties expected for the chemical kinetics equations, like positivity of entropy production. It cannot have bifurcations, oscillations, etc. The slow system in the QE approximation also has the thermodynamic properties.  This is a particular case of the general theorem about the {\em preservation of the  type of dynamics}: the QE approximation and the fast subsystem  inherit thermodynamic behaviour from the original system \cite{GorKarOtt2001}. This property makes QE a promising initial approximation for various iterative model reduction procedures,  \cite{GorKarOtt2001}.

 In the QSS, the fast system can have many steady states, oscillation, etc.  These critical effects in the  QSS approach are considered as the symptoms of physical critical effects, for example ignition: the original system with thermodynamic behaviour has no bifurcations but just high sensitivity, at the same time the fast system in  the QSS approximation has bifurcations.

If the fast system does not converge to equilibrium then the methods of {\em averaging} can work \cite{SlemrodAcharya2012}. Instead of partial equilibria or quasi-steady-states the so-called `ergodes' are used (subsets, on which the fast systems are ergodic).

All the methods of invariant manifold in chemical kinetics were introduced for generalisation and improvement of the QE and QSS.  The earliest prototype was the Chapman--Enskog method in the kinetic foundation of fluid dynamics but this method used the Taylor series expansion in powers of a small parameter. The second approximation from this series is already  singular. The iterative MIM was developed for the Boltzmann equations for correcting these singular properties of the Chapman--Enskog expansion \cite{GKTTSP94, GorbanKarlin2014} and for solution of  Hilbert's 6th Problem \cite{Slemrod2013}. The mathematically rigorous version of the Chapman--Enskog method for finite-dimensional ODE was developed  by Fenichel \cite{Fenichel1979} and was used as the prototype in many later works (see the collection of papers \cite{ModRed2006}).

\begin{figure}
\centering
\includegraphics[width=0.4\textwidth]{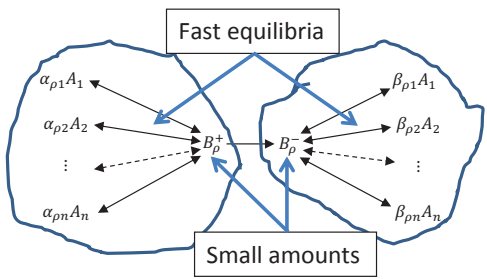}
\caption {The Michaelis--Menten--Stueckelberg limit: intermediates $B$ are in fast equilibrium with reagents $A_i$ and are present in small amount (QSS). This limit results in mass action law for the brutto reactions between $A_i$.}
\label{Fig:MMS_MAL}
\end{figure}

Combination of the QE and QSS assumptions (fast partial equilibria and small amounts of intermediates, for example) leads to important general results. It was used in 1913 by Michaelis and Menten. They proved that under this assumption the intermediates could be completely excluded from kinetic equations and the resulting kinetics will obey mass action law again. In 1952, Stueckelberg used this QE+QSS combination for proving the semi-detailed balance condition for Boltzmann's equations (later, this condition was rediscovered and is popularised as the {\em complex balance} condition). The detailed review was presented and the general Michaelis--Menten--Stueckelberg asymptotic theorem was formulated and proved in \cite{GorbanShah2011, GorbanKol2015} (see Fig.~\ref{Fig:MMS_MAL}).

If there exist no fast equilibria (no QE), then the QSS  approximation alone does not lead to the simple mass action law after exclusion of the intermediates. Nevertheless, if the mechanism of the intermediates transitions is linear then the explicit analytic expressions for the QSS reaction rates are obtained \cite{Yab1991}. The simplest of them, for the enzime (E) -- substrate (S) reaction, $S+E \leftrightarrows SE \to E+P$, is known as the Michaelis--Menten kinetics (produced by Briggs and Haldane in 1925).

 For non-linear reactions between intermediates, they usually cannot be excluded from the equations analytically but an equation for the reaction rate can be produced. The method of {\em kinetic polynomial} gives the algebraic equation for the reaction rate \cite{MarinYabl2011}. This equation (the kinetic polynomial) is produced by the exclusion methods (the constructive algebraic geometry) and is successfully used in analysis of catalytic reactions.

\begin{figure*}[t]
\centering
\includegraphics[width=0.7\textwidth]{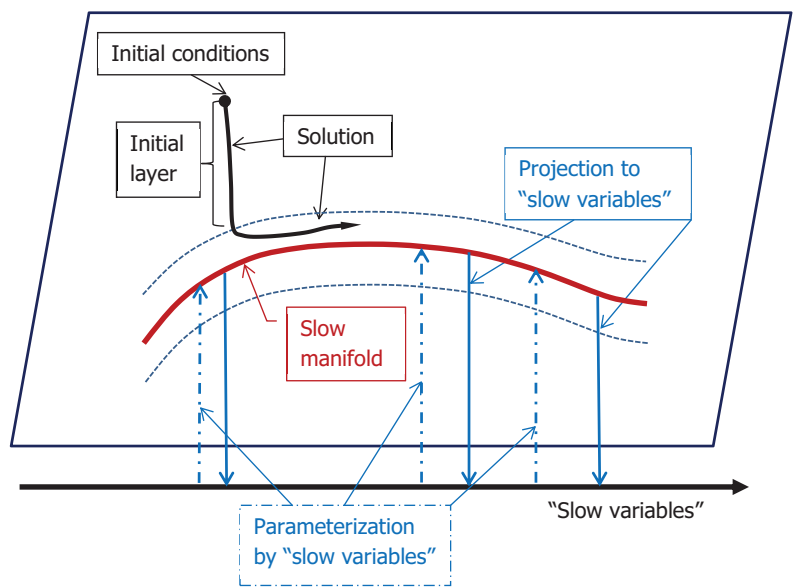}
\caption {Slow Invariant Manifold}
\label{Fig:MIM}
\end{figure*}

The MIM aims to find the slow invariant manifold (Fig.~\ref{Fig:MIM}) \cite{RouFra1990,  GKTTSP94, SinghPowers2002, GorbanKarlin2003}. Invariance has   simple definitions, both analytic and geometric ones. Consider a system of ODE, $\dot{x}=F(x)$.  Imagine that we found such a (nonlinear) coordinate transformation after  that the vector $x$ is represented as a direct sum $x=y\oplus z$ and the equations have the form 
\begin{equation}
\begin{split}
\dot{y}=F_y(y,z)\\
\dot{z}=F_{z}(y,z),
\end{split}
\end{equation}
where $F_{y}(0,z)=0$. This means that if $y(0)=0$ then $y(t)=0$ for solutions of the system.

The manifold $y=0$ is invariant with respect to the system. Of course, it is not necessary to assume a global transformation with this property. We can use local transformations under condition that they define the same manifolds $y=0$ in intersections (to provide gluing of the local pieces). The definition of slowness is much more sophisticated. It can be done rigorously for systems with existing slow parameters $\varepsilon$ using analytical continuations from the very small values of   $\varepsilon$ or from a vicinity of an attractor (for example, a stable fixed point) using separation of the relaxation modes into fast and slow ones (for example, separation of the invariant subspaces of Jacobian by real parts of eigenvalues) \cite{GorbanKarlin2014}. 

Far from these continuations, there exist two popular heuristics: one used the eigenspaces of Jacobians in various states \cite{GoussisMaas2011}, another relies on a special iteration procedures (a special version of the Newton--Krylov method) \cite{GorbanKarlinBook2005}. Both approaches   converge to a slow manifold for sufficiently small $\varepsilon$ or in a vicinity of an asymptotically stable equilibrium under some simple technical conditions (see \cite{GorbanKarlinBook2005, KapKapZag2015, WuKaper2017}).

A special model reduction method, which utilised the analytic continuation from a vicinity of a fixed point and Lyapunov auxiliary theorem was  developed \cite{KazantzisKravaris2010}.

The CSP  looks for more than just a slow manifold. It tries to find the complete slow-fast decomposition, that is to transform the initial system of ODE  $\dot{x}=F(x)$ into a decoupled form
\begin{equation}
\begin{split}
\dot{y}=F_y(y)\\
\dot{z}=F_{z}(y),
\end{split}
\end{equation}
where the non-diagonal terms (dependence of $\dot{y}$ on $z$ and $\dot{z}$ on $y$) vanish. Of course, this complete decomposition could be rarely found explicitly (as well as the slow invariant manifold) but the CSP tends to decrease the non-diagonal terms iteratively. Again, the definition of decomposition is simple and straightforward, but the attribution of slow/fast properties may be complicated and can be done rigorously for families of systems with a small parameter or near a stable attractor. The classical heuristic for the slow-fast CSP decomposition used the eigenvectors and eigenspaces of the Jacobians at all states but very recently one of the inventors of the CSP proposed a direct iteration method, which can lead to the same result avoiding expensive calculations of the egienvalues and eigenvectors in transient states \cite{LamNewCSP2018}.

During the last decade much work was done to improve the computational efficiency and extend the area of applicability of the MIM and CSP. In particular, a grid-based computational method for slow invariant manifolds was elaborated \cite{GorKarZin2004grids}.

The method of Reaction-Diﬀusion Manifolds (REDIM) \cite{BykovMaas2007,NEBykovMaas2017} was developed to extend the MIM for handling the diffusion processes in combustion problems \cite{NEBykovMaas2017}. It represents a modification of the Film Equation \cite{GorbanKarlinBook2005,CMIM2004} and of functional iteration technique \cite{RouFra1990} to approximate a slow invariant manifold for a reaction-diffusion system. Recently, the REDIM method was improved to be able to generate and to approximate a slow manifold of arbitrary dimension by employing its hierarchical structure \cite{NEBykovMaas2017}. The REDIM was applied for model reduction in various combustion systems . 

The CSP was generalised and implemented as a part of a general strategy for analysis and reduction of uncertain chemical kinetic models \cite{GalaValorani2017}.

Lumping analysis was reformulated as a particular case of MIM \cite{OkekeRoussel2015} and computational possibilities of this approach were tested. The reaction rate constants are usually defined with uncertainty. A Bayesian automated method was implemented for robust lumping for systems with parameter variability \cite{Dokoumetzidis2009}. The method works stepwise, reducing the system's dimension by one at each step.

The theory and methods of invariant manifolds are used for modern development of the theory of critical simplification. In 1944, Lev Landau noticed that near the loss of stability the
amplitude of the emergent ``principal motion'' satisfies a very simple equation. It is an example of the ``bifurcational parametric simplification''. In chemical kinetics, The concept of ``critical'' simplification was proposed in chemical kinetics by Yablonsky and Lazman (1996) for the oxidation of carbon monoxide over a platinum catalyst using a Langmuir-Hinshelwood mechanism. The main observation was a simplification of the mechanism at ignition and extinction points. This is a very general phenomenon known for various bifurcations.  For the equations of chemical kinetics, the theory of critical simplification was developed recently \cite{GolKrapYab2015} using the constructive theory of invariant manifolds.

Various methods of construction of slow invariant manifolds were tested and compared using a simple example \cite{ChiavazzoGorKar2007}.  Method of invariant grids \cite{GorKarZin2004grids} was employed in this work for iteratively solving the invariance equation. Various initial  approximations for the grid are considered such as QE, Spectral QE, ILDM and Symmetric Entropic ILDM (proposed earlier \cite{GorbanKarlin2003}). Slow invariant manifold was also computed
using CSP method. A comparison between method of invariant grids and CSP is also reported. 
Although CSP and the tested  MIM are based on completely different construction of iterations, the comparison shows very similar results in terms of accuracy of slow invariant manifold description.

Thermodynamics gives a convenient opportunity to formulate the QE assumption. If  the fast reactions obey thermodynamics then the corresponding thermodynamic potential is a Lyapunov function: it should change monotonically along the  fast motion \cite{Yab1991, GorbanKarlinBook2005, Hangos2010, Hangos2015}, and QE can be described as a conditional extremum (that is, maximum for entropies and free entropies or minimum  for free energies) of the thermodynamics potential. 
In this formalism the reactions are not needed, only the plane of fast motion is necessary. This simple idea \cite{GorbanKarlin2003} gave rise to many `constrained equilibria' approximations \cite{HiremathPope2013, RenAtAl2017}. For example, a new version of the QE was proposed, the spectral QE manifold, that consists of partial equilibria in the planes of fast motions, which are parallel to fast eigenspaces of Jacobian  at  equilibrium \cite{KooshkbaghiKarlin2016}.

Thermodynamics was used for comparing all reversible reactions by the same measure: entropy production.  It was proposed to exclude almost equilibrated reversible reactions from the reaction mechanism \cite{BykovYab1977}. The distance from the reaction equilibrium was measured by entropy production. The total entropy production should not change significantly in this procedure. (If we assume the detailed balance, then the entropy produced by a reversible reaction measures how far is it from equilibrium.) This method was used for defining of the compartments in the composition space with the different reaction mechanism for hydrogen combustion \cite{Dimitrov82}. 

Recently, this method was re-engineered and  applied to  auto-ignition of n-heptane with extraction of  two skeletal mechanisms,  and for analysis of spatially-varying premixed laminar flames \cite{KooshkbaghiKarlin2014}. For partially irreversible reactions, there exist special Lyapunov functions as well \cite{GorbanMirYab2013}. 

The fast-slow separation in the form  (\ref{QSS2}) may be hidden and should be revealed. Let us look on the system with fast-slow separation (Fig.~\ref{Fig:MIM}). Two domains can be distinguished there: the domain where the system change is slow and the domain where the change is relatively fast. Moreover, the fast area is expected to be larger (the system `shrinks' to the smaller-dimensional manifold of slow motions). This heuristic idea gave rise to development of a Singularly Perturbed Vector Fields (SPVF) theory \cite{Bykov2006} and of a Global Quasi-Linearization (GQL) approach in its framework \cite{Bykov2008}.

In this way, the transformation to the explicitly decomposed form  (\ref{QSS2}) is constructed by a linear interpolation of the system vector field (\ref{EqSlowExpl}) from a randomized sample of vectors $x$ \cite{Bykov2006,BykovGoldMa2008,BykovGrifSazh2013,Bykov2013}. The results of application of the GQL and the QSS were systematically compared \cite{YuBykov2018, Nave2017}. 

 The model reduction methods based on the randomised sampling and principal component analysis are systematically used under the name `Proper Orthogonal Decomposition'  (POD) \cite{Willcox2002}. The POD is  an a posteriori, data dependent projection method. The data
points are either given by samplings from experiments or by trajectories of
the physical system extracted from simulations of the full model (the so-called `snapshots').
This point of view is often translated into the
question: ``Find a subspace approximating a given set of data in an optimal least–
squares sense'' \cite{Pinnau2008}. Various linear and non-linear versions and generalisations of the principal component analysis are applied for dimensionality reductions of snapshots \cite{GorZin2010}. In particular, a modification of the POD with preservation of some original state variables is proposed \cite{Chu2011}.

 We can expect intensive development of new model reduction methods with applications of the classical and new {\em data mining techniques}: independent component analysis, various methods of clustering, artificial neural networks, and other machine learning methods. For example, the profile likelihood approach allows solving the model reduction problem together with the indetification problem and analysis of  parameter identifiability and designate likely candidates for reduction.  The following references demonstrate some other recent efforts in this direction \cite{GorbanZinovyev2010, HiremathPope2011, Amato2012, Perini2013, Mirgolbabaei2013, Zinovyev2015, LiYang2016}.

\begin{figure*}
\centering
 \boxed{a)\includegraphics[height=0.2\textwidth]{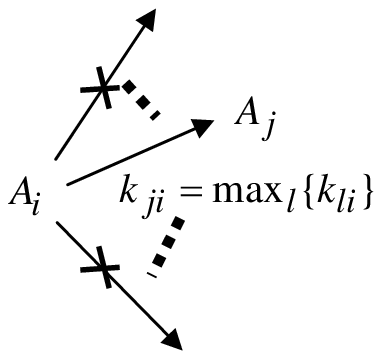}}\;
 \boxed{b) \includegraphics[height=0.2\textwidth]{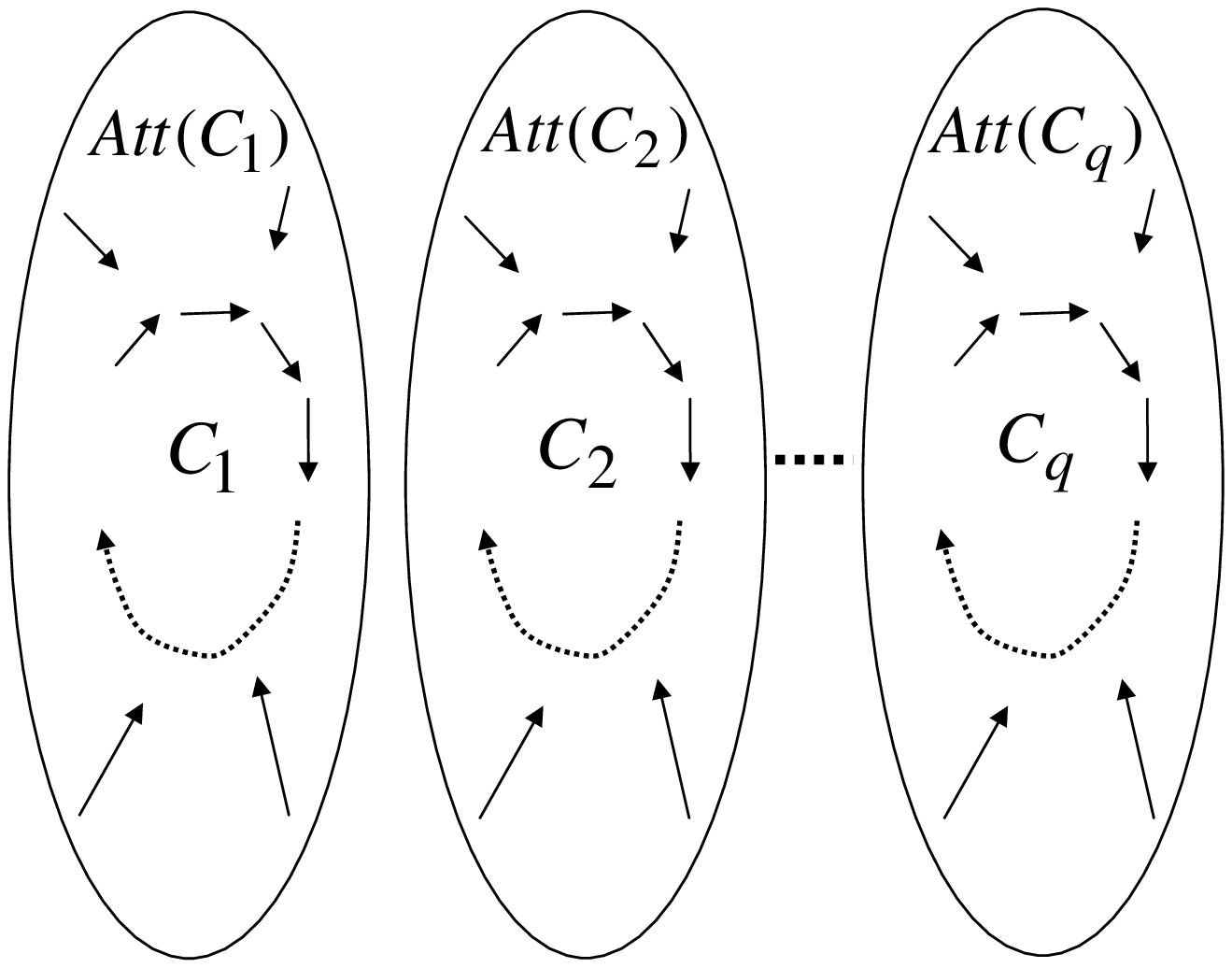}}\;
 \boxed{c) \includegraphics[height=0.2\textwidth]{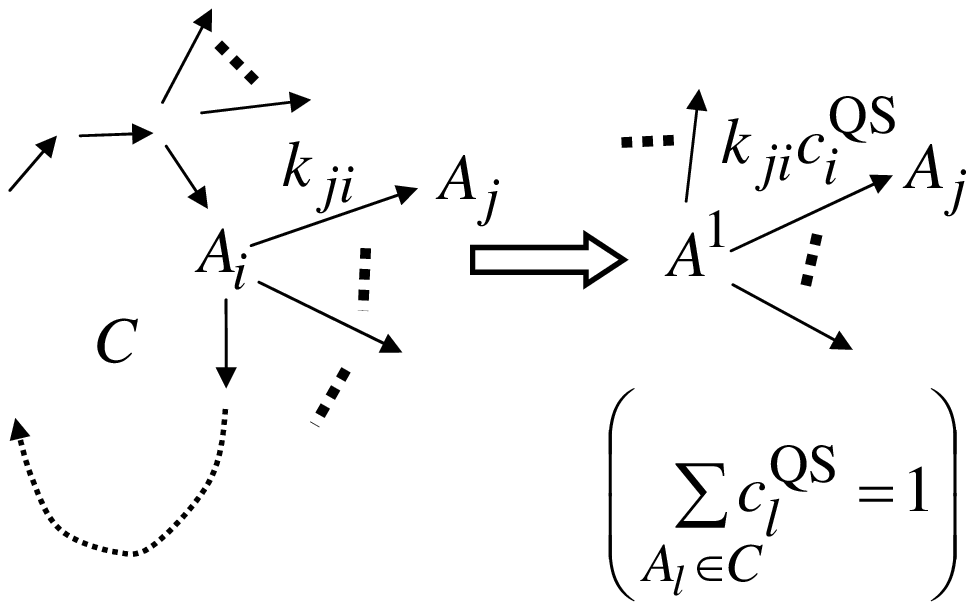}}
\caption {Hierarchical dominant path for linear system: (a) for each component retain the output reaction with maximal reaction rate:  in each reaction `fork' remains only dominant output and the reaction graph becomes a discrete dynamic system; (b) this discrete dynamic system converges to cycles (or fixed points); (c) cycles are glued in points, the output reactions are renormalised. A new reaction graph is constructed, where some vertices are glued. Then return to the step (a) for the new network. Iterate until all the trajectories of the discrete dynamical system converge to fixed points. This hierarchy of networks represents the hierarchical dominant path and provide us with asymptotic formulas for eigenvectors and eigenvalues of the initial network  \cite{GorbanRad2008,RadulescuBMC2008,GorRadZin2010}.}
\label{Fig:HierarchicalPath}
\end{figure*}

\section*{Reaction networks, limiting steps, and dominant paths}

Most of the methods described above can be applied to more general systems of ODE than reaction networks. They exploit the slow-fast separation  and some of them utilise the thermodynamic Lyapunov functions (like QE \cite{GorbanKarlin2003}). More rarely, the model reduction methods employ both the thermodynamic functions and reaction mechanism, like, for example, the method for extraction of skeletal mechanisms comparing entropy production by various reversible reactions \cite{KooshkbaghiKarlin2014}.

There exists a family of the model reduction methods, which intensively use the structure of reaction networks. The oldest, simplest, and, perhaps, the most used method of model reduction in chemical kinetics is the method of limiting step -- or bottleneck \cite{GorbanRad2008}.

Consider systems obeying mass action law: the reaction rate $r$ of the reaction $\sum_i \alpha_{\rho i} A_i \to \sum_i \beta_{\rho i} A_i$ is $r_{\rho}= k_{\rho} \prod_i c_i^{\alpha_i}$, where $A_i$ are the components, index $\rho$ is a reaction number,  $\alpha_{\rho i}$ and $\beta_{\rho i}$ are non-negative  stoichiometric coefficients, $k_{\rho}$ is the reaction rate constant, and $c_i$ is the concentration of $A_i$. The stoichiometric vector of the reaction $\gamma_{\rho}$ has coordinates $\gamma_{\rho i}=\beta_{\rho i}-\alpha_{\rho i}$. The kinetic equations (under constant volume) are
$$\dot{c}=\sum_{\rho} r_{\rho} \gamma_{\rho},$$
where $k$ is the reaction number.

A chain of  linear  irreversible reactions with non-zero rate constants $A_1\to A_2\to \ldots \to A_m$ has $m$ non-zero eigenvalues of the kinetic equation matrix, which  coincide with $-k_i$, where $k_i$ $(i=1,\ldots, m-1)$ is the reaction rate constant of the reaction $ A_i \to A_{i+1}$. Therefore, the relaxation rate is determined by the smallest constant. In a irreversible cycle  
$A_1 \to A_2 \to \ldots \to A_m \to A_1$ the smallest constant determines the stationary reaction rate. Indeed. in a steady state all the reaction rates in such a cycle should be equal: $k_i c_i = k_j c_j =w$. Therefore, in a steady state $c_i=w/k_j$, and $w=\sum_i c_i/\sum_j (1/k_j)$. If $0<k_1 < \varepsilon k_i$ ($i\neq 1$)  then $k_1 \sum_i c_i>w>k_1\sum_i c_i/(1+m\varepsilon)$. If  $0<\varepsilon\ll 1/m$ then $w \approx  k_1 \sum_i c_i$.  The relaxation time for an irreversible cycle with limiting step is inverse second reaction rate constant \cite{GorRadZin2010}. Indeed, the simple calculations show that for sufficiently small $\varepsilon$ the minimal non-zero eigenvalue of the kinetic matrix is close to the second constant in order.

The simple and widely used idea of the limiting steps was transformed into calculation of the hierarchical dominant paths (Fig.~\ref{Fig:HierarchicalPath})  for multiscale reaction networks \cite{GorRadZin2010}. Consider linear reaction network $A_i\to A_j$ with reaction rate constants $k_{ji}$.   Assume that in each reaction `fork' there exists a dominant reaction (Fig.~\ref{Fig:HierarchicalPath} a): if $k_{ji}=\max_l\{k_{li}\}$ then   $k_{qi}/k_{ji}< \varepsilon$ for some small parameter $\varepsilon$ and any $q\neq j$. Delete all the non-dominant reactions from the network. As a result, each component $A_i$ will have only one outgoing reaction. Thus,  the network is  transformed into a discrete dynamical systems, where $A_i$ are states and reactions are transitions. Every motion in such a dynamical system is attracted by a cycle or a fixed point (Fig.~\ref{Fig:HierarchicalPath} b). 

This is not the end of the story. Find for every  cycle $C$ the quasi stationary distribution: if we numerate the components in the cycle in the order of reactions $A_1,\ldots, A_k$ then $c_i^{\rm QS}= \frac{1}{k_i}/\sum_j\frac{1}{k_j}$, where $k_i$ is the reaction rate constant of the reaction $A_i \to \ldots$ from the cycle. Glue cycles into new vertices with the constants of outgoing reactions   multiplied by $c_i^{\rm QS}$ (Fig.~\ref{Fig:HierarchicalPath} c). Iterate the steps a-c of the analysis. The procedure will converge in the finite number of steps. We obtain a hierarchical dominant path: first, reaction converges to cycles (or fixed points), then, slower, to cycles of cycles, etc. It is proven \cite{GorRadZin2010} that the eigenvalues and eigenvectors of the initial network are well approximated by this procedure for sufficiently small $\varepsilon$. This approach was successfully applied to modelling the mechanisms of microRNA action \cite{Morozova2012, ZinovyevMorozova2013} and other biochemical reactions \cite{RadulescuBMC2008}

In order to find  the dominant subsystems and paths in nonlinear reaction networks, the methods of tropical (max,+) algebras were employed \cite{RadulescuVak2015, SamalGrig2015}.  The notion of tropical equilibration was elaborated that provides approximate descriptions of the slow invariant manifolds. Compared to computationally expensive numerical algorithms such as the CSP, the tropical approach is symbolic. It operates by the orders of magnitude instead of precise values of the model parameters.  The tropical methods provide  identification of metastable regimes, defined as the low dimensional regions of the phase space close to which the dynamics is much slower compared to the rest of the phase space. These metastable regimes depend on the network topology and on the orders of magnitude of the kinetic parameters.  

The local asymptotics of the algebraic curves are described by the tropical (or min-plus) systems of equations on their dominant exponents following the Newton polygon method. A necessary (tropical) condition of being a solution is the coincidence of the dominant orders of
growth of at least of two monomials of the equation. These conditions
provide a system of linear inequalities in the dominant exponents
which one can handle by means of the known algorithms for linear
programming. Thus, the problem of solving a system of algebraic
equations is reduced to a combinatorial one of determining the
dominant monomials. Since solving tropical systems is easier in general than solving
algebraic systems, one can get a benefit in calculating asymptotics.

 Extensive application of this method to   biochemical network models demonstrates that the number of dynamical variables in  the minimal models of large biochemical networks may be rather small and the  number of metastable regimes is sub-exponential in the number of variables and equations \cite{Grigoriev2016}.  The dynamics of the network can be described as a sequence of jumps from one metastable regime to another. A geometrically computed connectivity graph restricts the set of possible jumps. The graph theoretical symbolic preprocessing method significantly reduces computational complexity of the analysis of reaction networks and computation of parameter
regions for networks multistationarity  \cite{ENgland2017}.

Model reduction is a necessary tool for model composition in chemistry and biochemistry of large reaction networks \cite{KutumovaZin2013}. The model should be reduced even before its creation \cite{GorbanKarlin2003}. The full model can be imagined but its identification and reliable verification is often impossible. To obtain the model that works, we have to reduce the full model before identification. More precisely, model identification and model reduction should be combined in one process. There are several attempts of such an approach. For example,  the combination of model reduction and parameter estimation based on  Rao-Blackwellised particle filters decomposition methods was proposed for non-linear dynamical biochemical networks \cite{Sun2016} and tested successfully on synthetic and experimental data.

Asymptotology \cite{GorRadZin2010} and tropical asymptotics \cite{RadulescuVak2015, SamalGrig2015} operate by the orders of smallness rather than by computational accuracy and aim to find the limit `skeleton' of the reaction mechanism (under some additional conditions like preservation of connectivity or persistence).

Simultaneously, a new technique for model reduction was developed on the basis of the idea of controllable accuracy and error propagation \cite{Pepiot-Desjardins2008}. A geometric error propagation method creates a hierarchy of increasingly simplified kinetic schemes containing only important chemical paths. `Importance' is defined using preset accuracy targets and user-defined error-tolerance. The proposed technology operates with large and very large reaction mechanisms (hundreds and thousands of reactions). The model reduction in this technology is intrinsically connected to model identification and validation procedures: model reduction is an important component of these operations.

For example, the  advanced  Directed Relation Graph with Error Propagation and Sensitivity Analysis (DRGEPSA) method was developed \cite{Niemeyer2010}. Two skeletal mechanisms for n-decane were  generated by this method from a detailed reaction mechanism for n-alkanes. The detailed mechanism included  2115 species and 8157 reactions \cite{Niemeyer2010}.

Another reaction mechanism reduction method, Simulation Error Minimization Connectivity Method (SEM-CM) produces several consistent mechanisms, which include the preselected important species \cite{Nagy2009}. It starts from the   preselected important species and adds strongly connected sets of species. These strongly connected sets are identified using the Jacobian (sensitivities). This growth of the mechanism is controlled by the simulation accuracy and is terminated when the required accuracy is achieved. The method was tested on the reaction mechanism with  6874 reactions of 345 species. The reduced mechanism included 246 reactions of 47 species, and numerical simulations  became two orders faster \cite{Nagy2009}.

Various methods of reaction mechanism `surgery' is now one of the hottest topics in reduction of large and very large models in chemical engineering \cite{KooshkbaghiKarlin2014, Xin2014, Liu2016, Gao2016, Shakeri2017}. 
Surprisingly, the geometric details of this analysis seem to be very close to the geometric methods of asymptotology and tropical asymptotic.

\begin{figure*} 
\centering
\includegraphics[width=0.9\textwidth]{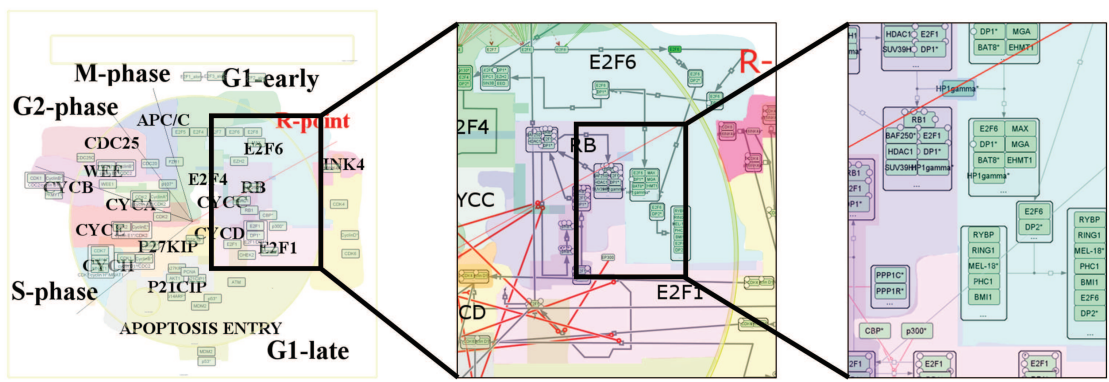}
\caption {Principle of semantic zooming exploited by NaviCell tool  \cite{Kuperstein2013} 
for visualizing complex reaction networks in cancer biology. Semantic zooming can be considering 
as a simple application of lumping for gradual hiding the details of complex molecular mechanisms.}
\label{Fig:NaviCell}
\end{figure*} 

For large and very large reaction systems, the reaction graph can be considered as an analogue of a geographic map of a large area. For such large graphs, lumping is close to zooming of maps. Zooming provides users by a tool for work on various levels of model granularity \cite{Sunnaker2011} and gives a possibility to study interaction between processes at different levels of the hierarchy. The principle of semantic zooming \cite{Kuperstein2013} was used for development  tools for navigations at different levels, similarly to geological information systems \cite{Dorel2017} (Fig.\ref{Fig:NaviCell}).

\section*{Conclusion}
Three components of model reduction methodology are proved to be useful: 
\begin{itemize}
\item Universal approaches developed for general finite-dimensional or even infinite dimensional systems (slow invariant manifolds, geometric singular perturbation theory, etc.). For them, chemical dynamics is an important source of  challenges and applications;
\item Thermodynamic approaches, which utilize the basic physical and chemical structures;
\item Algebraic approaches based on analysis of reaction mechanism.
\end{itemize}
Interaction and mutual enrichment of  these techniques will provide the future development of the  efficient model reduction for better computational performance and deeper understanding of chemical dynamics.

\section*{Acknowledgement}
{Supported by the University of Leicester, UK, and the Ministry of education and science of Russia (Project No. 14.Y26.31.0022).}


\begin{thebibliography}{10}

\bibitem{GorbanYablonsky2015}Gorban AN, Yablonsky GS: {\bf Three Waves of Chemical Dynamics.} {\em Math. Model. Nat. Phenom.} 2015, {\bf 10}(5):1--5.

\bibitem{GorbanYab2013}Gorban AN, Yablonsky GS: {\bf Grasping Complexity.} {\em Comput. Math. Appl.} 2013, {\bf 65}(10):1421--1426. 

\bibitem{RouFra1990}$\bullet$Roussel MR, Fraser SJ: {\bf Geometry of the steady--state
approximation: Perturbation and accelerated convergence methods.} {\em J. Chem. Phys.}  1990, {\bf 93}:1072--1081.

\bibitem{GKTTSP94}$\bullet$Gorban AN, Karlin IV: {\bf Method of invariant manifolds and
regularization of acoustic spectra.} {\em Transport Theory and Stat. Phys.} 1994, {\bf 23}:559--632.

\bibitem{Maas1992}$\bullet$Maas U, Pope SB: {\bf Simplifying chemical kinetics: intrinsic low--
dimensional manifolds in composition space.} {\em Combust. Flame} 1992, {\bf 88}, 239--264.

\bibitem{Lam1994}$\bullet$Lam  SH, Goussis DA: {\bf The CSP Method for Simplifying Kinetics.}
{\em International Journal of Chemical Kinetics} 1994 {\bf 26}, 461--486.

\bibitem{RabiSens1983}$\bullet$Rabitz H, Kramer M, Dacol D: {\bf Sensitivity analysis in chemical
kinetics.} {\em Ann. Rev. Phys. Chem.} 1983, {\bf 34}:419--461.

\bibitem{LiRabitz1990}$\bullet$Li G, Rabitz H: {\bf A general analysis of approximate lumping in chemical kinetics.} {\em Chemical Engineering Science} 1990, {\bf 45}(4):977--1002.

\bibitem{Ranzietal2008}Ranzi E, Cuoci A, Faravelli T, Frassoldati A, Migliavacca G, Pierucci S, Sommariva S: {\bf Chemical kinetics of biomass pyrolysis} {\em Energy  Fuels} 2008, {\bf 22}(6):4292--4300.

\bibitem{Nguyenetal2017}Nguyen TT, Teratani S, Tanaka R, Endo A, Hirao M: {\bf Development of a Structure-Based Lumping Kinetic Model for Light Gas Oil Hydrodesulfurization.} {\em Energy  Fuels} 2017 {\bf 31}(5):5673--5681.

\bibitem{Abou-Jaoude2016}Abou-Jaoud{\'e} W, Thieffry D, Feret J: {\bf Formal derivation of qualitative dynamical models from biochemical networks.} {\em Biosystems} 2016 {\bf 149}:70--112.

\bibitem{Newman2011}$\bullet$Newman M, Barabasi AL, Watts DJ: {\em The Structure and Dynamics of Networks.} Princeton University Press, Princeton, NJ; 2011.

\bibitem{GorbanKarlinBook2005}Gorban AN, Karlin IV. {\em Invariant Manifolds for Physical and Chemical Kinetics}. Springer, Berlin -- Heidelberg; 2005.
 
\bibitem{CMIM2004}Gorban AN, Karlin IV, Zinovyev AY: {\bf Constructive methods of invariant manifolds for kinetic problems.} {\em Phys. Reports} 2004, {\bf 396}:197--403.

\bibitem{TuranyiTomlin2014}$\bullet$Tur{\'a}nyi T, Tomlin AS:  {\em Analysis of Kinetic Reaction Mechanisms}. Springer, Berlin -- Heidelberg;  2014.

\bibitem{GoussisMaas2011}Goussis DA, Maas U: {\bf Model reduction for combustion chemistry.}   In {\em Turbulent Combustion Modeling}. Springer, Berlin -- Heidelberg 2011:193--220.

\bibitem{GearKaper2005}$\bullet$Gear CW, Kaper TJ, Kevrekidis IG, Zagaris A: {\bf Projecting to a slow manifold: Singularly perturbed systems and legacy codes.} {\em SIAM Journal on Applied Dynamical Systems} 2005, {\bf 4}(3):711-732.

\bibitem{UNIMOLD2004}Gorban AN, Karlin IV: {\bf Uniqueness of thermodynamic projector and kinetic basis of molecular individualism.} {\em Physica A} 2004, {\bf 336}(3-4):391--432.  

\bibitem{GorbanKarlin2003}$\bullet$Gorban AN, Karlin IV: {\bf Method of invariant manifold for chemical kinetics.} {\em Chemical Engineering Science} 2003, {\bf 58}(21):4751-4768


\bibitem{SegelSlemrod1989}Segel LA, Slemrod M: {\bf The quasi-steady-state assumption: A case study in perturbation.} {\em SIAM Rev.} 1989, {\bf 31}:446--477.

\bibitem{Yab1991}Yablonskii GS, Bykov VI, Gorban AN, Elokhin VI: {\em  Kinetic
Models of Catalytic Reactions.}  Elsevier, Amsterdam 1991.

\bibitem{LeeOthmer2010}Lee CH, Othmer HG: {\bf A multi-time-scale analysis of chemical reaction networks: I. Deterministic systems.} {\em J. Math. Biol.} 2010 {\bf 60}(3):387--450.

\bibitem{WestBiktashev2015}West S, Bridge LJ, White MR, Paszek P, Biktashev VN: {\bf A method of ‘speed coefficients’ for biochemical model reduction applied to the NF-$\kappa $B system.} {\em J. Math. Biol.} 2015 {\bf 70}(3):591--620.

\bibitem{GorKarOtt2001}Gorban AN, Karlin IV, Ilg P, {\"O}ttinger HC:
{\bf Corrections and enhancements of quasi-equilibrium states.} {\em J. Non-Newtonian Fluid Mech.} 2001, {\bf 96}:203--219.

\bibitem{GorBykYab1986}Gorban AN, Bykov VI, Yablonskii GS:
{\bf Thermodynamic function analogue for reactions proceeding without interaction of various substances.} {\em Chem. Eng. Sci.} 1986, {\bf 41}(11):2739--2745.

\bibitem{GorbanKarlin2014}$\bullet$$\bullet$Gorban AN, Karlin I: {\bf Hilbert's 6th Problem: exact and approximate hydrodynamic manifolds for kinetic equations.} {\em Bulletin of the American Mathematical Society} 2014, {\bf 51}(2):187--246.

Method of slow invariant manifolds for kinetic equations is explained in detail with applications to Boltzmann's equation.

\bibitem{Slemrod2013}Slemrod M: {\bf From Boltzmann to Euler: Hilbert's 6th problem revisited.} {\em Comput. Math. Appl.} 2013 {\bf 65}(10):1497--1501.

\bibitem{SlemrodAcharya2012}Slemrod M, Acharya A: {\bf Time-averaged coarse variables for multi-scale dynamics.} {\bf Quarterly of applied mathematics} 2012, {\bf 70}(4):793--803.
 
\bibitem{GorbanShah2011}$\bullet$Gorban AN, Shahzad M: {\bf The Michaelis-Menten-Stueckelberg theorem.} {\em Entropy} 2011, {\bf 13}(5):966--1019.

\bibitem{GorbanKol2015}Gorban AN, Kolokoltsov VN: {\bf Generalized mass action law and thermodynamics of nonlinear Markov processes.} {\em Math. Model. Nat. Phenom.} 2015, {\bf 10}(5):16--46.


\bibitem{MarinYabl2011}$\bullet$Marin G, Yablonsky GS: {\bf Kinetics of chemical reactions. Decoding complexity.} John Wiley \& Sons, Weincheim, 201c1.

A modern textbook in chemical kinetics, experimental and theoretical, fundamental and applied. In particular, in Chapter 9 the theory of kinetic polynomials is presented. 

\bibitem{Fenichel1979}Fenichel N: {\bf Geometric singular perturbation theory for ordinary differential equations.} {\em J. Diff. Eq.} 1979, {\bf 31}:59--93.

\bibitem{ModRed2006} Gorban AN,  Kazantzis N,  Kevrekidis IG,  {\"O}ttinger HC, Theodoropoulos C (Eds.): {\em Model Reduction and Coarse--Graining Approaches for Multiscale Phenomena.}
Springer, Berlin-Heidelberg-New York 2006.




\bibitem{SinghPowers2002}Singh S, Powers JM, Paolucci S. {\bf On slow manifolds of chemically reactive systems.} {\em J. Chem. Phys.} 2002, {\bf 117}(4):1482--1496.

\bibitem{KapKapZag2015}Kaper HG, Kaper TJ, Zagaris A: {\bf Geometry of the computational singular perturbation method.} {\em Math. Model. Nat. Phenom.} 2015, {\bf 10}(3):16--30.

\bibitem{WuKaper2017}$\bullet$Wu X, Kaper TJ: {\bf Analysis of the approximate slow invariant manifold method for reactive flow equations.} {\em Journal of Mathematical Chemistry} 2017, {\bf 55}(9):1--30.

Approximate Slow Invariant Manifold method is carefully analysed fo   reaction–diffusion equations with slow and fast reaction kinetics. The second-order errors were precisely evaluated. The results were also illustrated with a simple benchmark model.  

\bibitem{KazantzisKravaris2010}Kazantzis N, Kravaris C, Syrou L: {\bf A new model reduction method for nonlinear dynamical systems.} {\em Nonlinear Dyn.} 2010 {\bf 59}(1):183--194.

\bibitem{GOussis2013}Goussis DA: {\bf Quasi steady state and partial equilibrium approximations: their relation and their validity.} {\em Combust. Theor. Model.} 2012, {\bf 16}(5):869--926.

\bibitem{GOussis2015}Goussis DA. {\bf Model reduction: When singular perturbation analysis simplifies to partial equilibrium approximation.} {\em Combust. Flame} 2015, {\bf 162}(4):1009--1018.


\bibitem{LamNewCSP2018}$\bullet$ $\bullet$ Lam SH: {\bf 
An Efficient Implementation of Computational Singular Perturbation.} {\em Combustion Science and Technology} 2018, {\bf 190}(1):157--163.

An efficient formulation of computational singular perturbation without calculation of Jacobian eigenvectors and eigenvalues is proposed and tested on simple example.

\bibitem{GorKarZin2004grids}Gorban AN, Karlin IV, Zinovyev AY: {\bf Invariant grids for reaction kinetics.} {\em Physica A} 2004 {\bf 333}:106-54.


\bibitem{BykovMaas2007}Bykov V, Maas U: {\bf The extension of the ILDM concept to reaction-diffusion manifolds.} {\em Combust. Theor. Model.} 2007 {\bf 11}(6):839--862.

\bibitem{NEBykovMaas2017}$\bullet$Neagos A, Bykov V, Maas U: {\bf Adaptive hierarchical construction of reaction--diffusion manifolds for simplified chemical kinetics.} {\em Proceedings of the Combustion Institute} 2017 {\bf 36}(1):663--672.

The concept of slow manifold was successfully implemented for reaction-diffusion equations. The method of  Reaction--Diffusion Invariant Manifolds (REDIM) was developed

\bibitem{StrassackerBykov2018}Strassacker C, Bykov V, Maas U: {\bf REDIM reduced modeling of quenching at a cold wall including heterogeneous wall reactions.} {\em Int. J. Heat Fluid Flow} 2018, {\bf 69}:185--193.

\bibitem{GalaValorani2017}$\bullet$Galassia RM, Valorani M,  Najm HN, Safta C,  Khalil M, Ciottoli PP: {\bf Chemical model reduction under uncertainty.} {\em Combust. Flame} 2017,  {\bf 179}:242--252

CSP is  employed to generate simplified kinetic mechanisms, starting from a detailed reference mechanism with uncertainty. Uncertainty is modelled by random rate constants.

\bibitem{OkekeRoussel2015}Okeke BE, Roussel MR: {\bf An Invariant-Manifold Approach to Lumping.} {\em Math. Model. Nat. Phenom.} 2015, {\bf 10}(3):149--67


\bibitem{Dokoumetzidis2009}Dokoumetzidis A, Aarons L: {\bf A method for robust model order reduction in pharmacokinetics.} {\em Journal of pharmacokinetics and pharmacodynamics} 2009, {\bf 36}(6):613.

\bibitem{GolKrapYab2015}$\bullet$Gol'dshtein V,  Krapivnik N,  Yablonsky GS: {\bf About bifurcational parametric simplification.} {\em Math. Model. Nat. Phenom.} 2015, {\bf 10}(3):168--185.


\bibitem{ChiavazzoGorKar2007}$\bullet$Chiavazzo E, Gorban AN, Karlin IV. {\bf Comparison of invariant manifolds for model reduction in chemical kinetics.} {\em  Commun. Comput. Phys.} 2007, {\bf 2}(5):964-92.

\bibitem{RenAtAl2017}$\bullet$Ren Z, Lu Z, Gao Y, Lu T, Hou L: {\bf A kinetics-based method for constraint selection in rate-controlled constrained equilibrium.} {\em Combust. Theor. Model.} 2017,  {\bf 21}(2):159--182.


The rate-controlled constrained-equilibrium (RCCE) method is compared to the classical QE approximation and it is demonstrated that for the systems with slow and fast reactions these methods are equivalent The detailed analysis allows authors to claim that the thermodynamics-based constrained equilibrium manifolds (CEMs) give a good approximation to the actual  slow invariant manifolds   in reactive systems. 

\bibitem{Hangos2010}Hangos KM: {\bf Engineering model reduction and entropy-based Lyapunov functions in chemical reaction kinetics.} {\em  Entropy} 2010, {\bf 12}(4):772--797.

\bibitem{HiremathPope2013}$\bullet$Hiremath V, Pope SB: {\bf A study of the rate-controlled constrained-equilibrium dimension reduction method and its different implementations.} {\em Combust. Theor. Model.} 2013,  {\bf 17}(2):260--293.

\bibitem{Hangos2015}Hangos KM, Magyar A, Szederk{\'e}nyi G: {\bf Entropy-inspired Lyapunov functions and linear first integrals for positive polynomial systems.} {\em Math. Model. Nat. Phenom.} 2015, {\bf 10}(3):105--123.

\bibitem{KooshkbaghiKarlin2016}Kooshkbaghi M, Frouzakis CE, Boulouchos K, Karlin IV: {\bf Spectral quasi-equilibrium manifold for chemical kinetics.} {\em The Journal of Physical Chemistry A.} 2016 {\bf 120}(20):3406-3413.


\bibitem{BykovYab1977}Bykov VI,  Yablonskii GS, Akramov TA, Slin'ko MG: {\bf The rate of the free energy decrease in the course of the complex chemical reaction.} {\em  Dokl. Akad. Nauk USSR} 1977, {\bf 234}(3):621--624.
 
\bibitem{Dimitrov82}Dimitrov VI: {\em Simple kinetics}, Nauka, Novosibirsk, 1982.

\bibitem{KooshkbaghiKarlin2014}$\bullet$Kooshkbaghi M, Frouzakis CE, Boulouchos K, Karlin IV: {\bf Entropy production analysis for mechanism reduction.} {\em  Combust. Flame} 2014, {\bf 161}(6):1507-1515.

A method is developed for eliminating species from detailed reaction mechanisms in order to generate skeletal schemes. The  approach is based on the relative contribution of each elementary reaction to the total entropy production.

\bibitem{GorbanMirYab2013}$\bullet$Gorban AN, Mirkes EM, Yablonsky GS: {\bf Thermodynamics in the limit of irreversible reactions.} {\em Physica A}  2013, {\bf 392}(6):1318--1335.

 The systems with irreversible reactions are represented as the limits of fully reversible systems when some of the equilibrium concentrations tend to zero.   If the reversible systems obey the principle of detailed balance then the limit system with some irreversible reactions must satisfy the extended principle of detailed balance:

(i) The reversible part satisfies the principle of detailed balance; 

(ii) The convex hull of the stoichiometric vectors of the irreversible reactions does not intersect the linear span of the stoichiometric vectors of the reversible reactions. 
 
These conditions imply the existence of the global Lyapunov functionals. Thermodynamic theory of the irreversible limit of reversible reactions is illustrated by the analysis of hydrogen combustion.


\bibitem{Bykov2006}Bykov V, Goldfarb I, Gol'dshtein V: {\bf Singularly Perturbed Vector Fields,} {\em J. Phys.: Conf. Ser.} 2006, {\bf 55}:28--44.

\bibitem{Bykov2008}Bykov V, Gol'dshtein V: {\bf On a Decomposition of Motions and Model Reduction.} {\em J. Phys.: Conf. Ser.} 2008, {\bf 138}:012003.

\bibitem{BykovGoldMa2008}$\bullet$Bykov V, Gol'dshtein V, Maas U: {\bf Simple global reduction technique based on decomposition approach.} {\em Combust. Theor. Model.} 2008, {\bf 12}(2):89--405.

\bibitem{BykovGrifSazh2013}Bykov V, Griffiths JF, Piazesi R, Sazhin SS, Sazhina EM: {\bf The application of the Global Quasi-Linearisation technique to the analysis of the cyclohexane/air mixture autoignition.} {\em Applied Mathematics and Computation} 2013, {\bf 219}(14):7338--7347.

\bibitem{YuBykov2018}$\bullet$Yu C, Bykov V, Maas U. {\bf Global Quasi-Linearization (GQL) versus QSSA for a hydrogen-air auto-ignition problem.} {\em 	Phys. Chem. Chem. Phys.} 2018, DOI:10.1039/C7CP07213A.

Global quasi-linearization was explained and systematically compared with QSS.

\bibitem{Nave2017}Nave O: {\bf Singularly Perturbed Vector Field Method (SPVF) Applied to Combustion of Monodisperse Fuel Spray.} {\em Differential Equations and Dynamical Systems} 2017:1--18.


\bibitem{Bykov2013}Bykov V, Gol'dshtein V: {\bf Fast and Slow Invariant Manifolds for Chemical Kinetics.}{\em  Comput. Math. Appl.}  2013 {\bf 65}(10), 1502--1515.

\bibitem{Willcox2002}Willcox K, Peraire J: {\bf Balanced model reduction via the proper orthogonal decomposition.} {\em AIAA journal} 2002, {\bf 40}(11):2323-30.

\bibitem{Pinnau2008}Pinnau R: {\bf Model reduction via proper orthogonal decomposition.} In {\em Model Order Reduction: Theory, Research Aspects and Applications Mathematics in Industry book series (MATHINDUSTRY, volume 13).} Springer, Berlin -- Heidelberg 2008:95--109.

\bibitem{GorZin2010} Gorban AN, Zinovyev A: {\bf Principal manifolds and graphs in practice: from molecular biology to dynamical systems.} {\em Int. J. Neural Syst.} 2010, {\bf 20}(3):219--232.

\bibitem{Chu2011}Chu Y, Serpas M, Hahn J: {\bf State-preserving nonlinear model reduction procedure.} {\em Chem. Eng. Sci.} 2011 {\bf 66}(17):3907--3913.

\bibitem{Maiwald2016}Maiwald T, Hass H, Steiert B, Vanlier J, Engesser R, Raue A, Kipkeew F, Bock HH, Kaschek D, Kreutz C, Timmer J: {\bf Driving the Model to Its Limit: Profile Likelihood Based Model Reduction.} {\em PloS One} 2016,  {\bf 11}(9):e0162366.

\bibitem{GorbanZinovyev2010}Gorban AN, Zinovyev A {\bf Principal manifolds and graphs in practice: from molecular biology to dynamical systems.} {\em International journal of neural systems} 2010, {\bf 20}(03):219--232.

\bibitem{HiremathPope2011}Hiremath V, Ren Z, Pope SB: {\bf Combined dimension reduction and tabulation strategy using ISAT–RCCE–GALI for the efficient implementation of combustion chemistry.}  {\em Combust. Flame} 2011, {\bf 158}(11):2113--2127.

\bibitem{Amato2012}Amato F, Gonz{\'a}lez-Hern{\'a}ndez JL, Havel J: {\bf Artificial neural networks combined with experimental design: a ``soft'' approach for chemical kinetics.} {\em Talanta} 2012, {\bf 93}:72--78.

\bibitem{Perini2013}Perini F: {\bf High-dimensional, unsupervised cell clustering for computationally efficient engine simulations with detailed combustion chemistry.} {\em Fuel} 2013, {\bf 106}:344--356.

\bibitem{Mirgolbabaei2013}Mirgolbabaei H, Echekki T: {\bf A novel principal component analysis-based acceleration scheme for LES–ODT: An a priori study.} {\em Combust. Flame} 2013, {\bf 160}(5):898--908.

\bibitem{Zinovyev2015}$\bullet$Zinovyev A: {\bf Overcoming Complexity of Biological Systems: from Data Analysis to Mathematical Modeling.} {\em  Math. Model. Nat. Phenom.} 2015, {\bf 10}(3):186-205.

A detailed review about complexity reduction in biochemical systems.

\bibitem{LiYang2016}Li S, Yang B, Qi F:  {\bf Accelerate global sensitivity analysis using artificial neural network algorithm: Case studies for combustion kinetic model.} {\em Combust. Flame} 2016,  {\bf 168}:53--64.



\bibitem{GorbanRad2008}Gorban AN, Radulescu O: {\bf Dynamic and static limitation in reaction networks, revisited.} {\em Adv. Chem. Eng.} 2008, {\bf 34}103--173.

\bibitem{GorRadZin2010}$\bullet$ $\bullet$Gorban AN, Radulescu O, Zinovyev AY: {\bf Asymptotology of chemical reaction networks.} {\em Chem. Eng. Sci.} 2010, {\bf 65}:2310--2324.

\bibitem{Morozova2012}Morozova N, Zinovyev A, Nonne N, Pritchard LL, Gorban AN, Harel-Bellan A: {\bf Kinetic signatures of microRNA modes of action.} {\em RNA} 2012, {\bf 18}(9):1635--1655.

\bibitem{ZinovyevMorozova2013}$\bullet$Zinovyev A, Morozova N, Gorban AN,   Harel-Belan A: {\bf Mathematical Modeling of microRNA-Mediated Mechanisms of Translation Repression.} In {\em  MicroRNA Cancer Regulation: Advanced Concepts, Bioinformatics and Systems Biology Tools}, Advances in Experimental Medicine and Biology series,  Vol. 774, Springer, 2013:189--224.

\bibitem{RadulescuBMC2008}Radulescu O, Gorban AN, Zinovyev A, Lilienbaum A: {\bf Robust simplifications of multiscale biochemical networks.} {\em BMC Syst. Biol.	} 2008, {\bf 2}(1):86.
.
\bibitem{RadulescuVak2015}$\bullet$Radulescu O, Vakulenko S, Grigoriev D: {\bf Model Reduction of Biochemical Reactions Networks by Tropical Analysis Methods.} {\em Math. Model. Nat. Phenom.} 2015, {\bf 10}(3):124--138.

Theory of tropical asymptotics (max,+) algebras and geometry of tropical equilibration of chemical reactions.


\bibitem{SamalGrig2015}Samal SS, Grigoriev D, Fröhlich H, Weber A, Radulescu O: {\bf A geometric method for model reduction of biochemical networks with polynomial rate functions.} {\em Bull. Math. Biol.} 2015, {\bf 77}(12):210-2211.

\bibitem{Grigoriev2016}Grigoriev D, Radulescu O, Samal S, Naldi A, Theret N, Weber A: {\bf A
Geometric analysis of pathways dynamics: application to versality of
TGF-beta receptors.}  {\em Biosystems} 2016, {\bf 149}:3--14.

\bibitem{ENgland2017}England M, Errami H, Grigoriev D, Radulescu O, Sturm T, Weber A:
{\bf Symbolic versus numerical computation and vizualization of parameter
regions for multistationarity of biological networks.}  {\em Proc. Intern.
Symp. Comp. Algebr. Symb. Comput., Beijing, 2017, Lect. Notes Comput.
Sci.} vol. 10490:93--108.


\bibitem{KutumovaZin2013}Kutumova E, Zinovyev A, Sharipov R, Kolpakov F. {\bf Model composition through model reduction: a combined model of CD95 and NF-κB signaling pathways.} {\em  BMC Syst. Biol.	} 2013, {\bf 7}(1):13.

\bibitem{Sun2016}Sun X, Medvedovic M: {\bf Model reduction and parameter estimation of non-linear dynamical biochemical reaction networks.} {\em IET Syst. Biol.	} 2016, {\bf 10}(1):10-16.

\bibitem{Pepiot-Desjardins2008}Pepiot-Desjardins P, Pitsch H: {\bf An efficient error-propagation-based reduction method for large chemical kinetic mechanisms.} {\em Combust. Flame} 2008, {\bf 154}(1):67-81.

\bibitem{Niemeyer2010}$\bullet$Niemeyer KE, Sung CJ, Raju MP: {\bf Skeletal mechanism generation for surrogate fuels using directed relation graph with error propagation and sensitivity analysis.}  {\em Combust. Flame} 2010, {\bf 157}(9):1760--1770.

\bibitem{Nagy2009}Nagy T, Tur{\'a}nyi T: {\bf Reduction of very large reaction mechanisms using methods based on simulation error minimization.} {\em Combust. Flame} 2009, {\bf 156}(2):417--428.

\bibitem{Xin2014}Xin Y, Sheen DA, Wang H, Law CK: {\bf Skeletal reaction model generation, uncertainty quantification and minimization: Combustion of butane.} {\em Combust. Flame} 2014, {\bf 161}(12):3031--3039.

\bibitem{Liu2016}Liu T, Jiaqiang E, Yang W, Hui A, Cai H: {\bf Development of a skeletal mechanism for biodiesel blend surrogates with varying fatty acid methyl esters proportion.} {\em Appl. Energy} 2016, {\bf 162}:278--288.

\bibitem{Gao2016}Gao X, Yang S, Sun W: {\bf A global pathway selection algorithm for the reduction of detailed chemical kinetic mechanisms.} {\em Combust. Flame} 2016, {\bf 167}:238--247.

\bibitem{Shakeri2017}Shakeri A, Mazaheri K, Owliya M: {\bf Using sensitivity analysis and gradual evaluation of ignition delay error to produce accurate low-cost skeletal mechanisms for oxidation of hydrocarbon fuels under high-temperature conditions.} {\em Energy  Fuels} 2017, {\bf 31}(10):11234--11252.

\bibitem{Sunnaker2011}Sunn\r{a}ker M, Cedersund G, Jirstrand M: {\bf A method for zooming of nonlinear models of biochemical systems.} {\em BMC Syst. Biol	} 2011 {\bf 5}(1):140.

\bibitem{Kuperstein2013}$\bullet$Kuperstein I, Cohen DP, Pook S, Viara E, Calzone L, Barillot E, Zinovyev A: {\bf NaviCell: a web-based environment for navigation, curation and maintenance of large molecular interaction maps.} {\em BMC Syst. Biol	} 2013 {\bf 7}(1):100.

NaviCell Web Service (\url{http://navicell.curie.fr})  is a tool for network-based visualization of `omics' data which implements several data visual representation methods and utilities for combining them together. It uses Google Maps and semantic zooming to browse large biological network maps with different types of the molecular data mapped on top of them. 

\bibitem{Dorel2017}$\bullet$Dorel M, Viara E, Barillot E, Zinovyev A, Kuperstein I: {\bf NaviCom: a web application to create interactive molecular network portraits using multi-level omics data.} {\em Database} 2017,  {\em 2017}(1):bax026.  

NaviCom, a Python package and web platform for visualization of multi-level omics data on top of biological network maps. NaviCom is bridging the gap between cBioPortal, the most used resource of large-scale cancer omics data and NaviCell, a data visualization web service that contains several molecular network map collections. 

\end{thebibliography}
\end{document}